\documentclass[prl,superscriptaddress,showpacs,longbibliography,reprint]{revtex4-2} 

\usepackage[colorlinks=true,linkcolor=blue,urlcolor=blue,citecolor=blue,pdfusetitle]{hyperref}
\usepackage{color}
\usepackage[utf8]{inputenc}
\usepackage[usenames,dvipsnames]{xcolor}
\usepackage{amsmath}
\usepackage{amssymb}
\usepackage{graphicx}
\graphicspath{{graphics/}}
\usepackage{epsfig}
\usepackage{dcolumn}
\usepackage{soul}
\usepackage{bm}
\usepackage{mathrsfs}
\usepackage{multirow}
\usepackage[all]{xy}
\usepackage{pbox}
\usepackage{lipsum}
\usepackage{verbatim}
\usepackage{braket}
\usepackage{dsfont}
\usepackage{array}
\usepackage{makecell}
\usepackage{tabularx}
\usepackage{isotope}
\usepackage{xr}
\usepackage{float}

\usepackage{ntheorem}

\newtheorem*{theorem-non}{Theorem.}
\newtheorem*{lemma-non}{Lemma.}
\newtheorem*{corollary-non}{Corollary.}

\usepackage{todonotes}
\usepackage{mathtools}

\begin{document}
\raggedbottom

\title{Don't truncate, decompose: mean-field dynamics of long-range quantum systems \\from strongly correlated states}

\author{Federico Carollo}
\affiliation{Dipartimento di Fisica, Sapienza Università di Roma, Piazzale Aldo Moro 5, 00185 Rome, Italy}

\date{\today}

\begin{abstract}
We challenge the widespread consensus that mean-field theory fails to describe long-range open quantum systems in the presence of symmetry breaking and/or when starting from strongly correlated states (e.g., macroscopic superpositions). While recent literature relies on cumulant expansions to capture such systems, this approach rests  on truncations with no clear justification. Here, we show that it is, at best, conceptually redundant in the strong long-range regime. We show that the evolution can be decomposed into, and fully  reconstructed from, independent mean-field dynamics. This decomposition generates the entire hierarchy of cumulants and, as a byproduct, identifies---to our knowledge, for the first time---a regime in which cumulant expansions exactly predict low-order cumulants. We illustrate the power of our findings with two applications: we compute the moment generating function for nonequilibrium $\mathcal{Z}_2$ symmetry breaking, and construct states restoring time-translation symmetry  in time crystals. In both cases, our method fully reproduces the exact many-body dynamics, which is out of reach of cumulant expansions. Our results reclaim the exactness of mean-field theory, offering a transparent framework for large-scale open quantum systems.
\end{abstract}

\maketitle

Long-range interacting systems have emerged as a vibrant research direction in quantum science \cite{defenu2023,defenu2024}. Compared to their short-range counterparts, they enable fundamentally new phenomena, including quasi-stationary states \cite{kastner2011,defenu2021} and anomalous information propagation \cite{eisert2013,richerme2014,jurcevic2014}.  Long-range interactions are routinely realized across a diverse set of  architectures, from trapped ions and Rydberg-atom arrays \cite{saffman2010,browaeys2020,monroe2021,steinert2023,chen2023,foss-feig2025,bai2026} to ensemble of atoms in optical cavities or in free space \cite{vaidya2018,mivehvar2021,ferioli2023,bonifacio2024,cooper2024,bohr2024}. In these platforms, typically described within the framework of open quantum systems \cite{breuer2002,gardiner2004} due to unavoidable dissipative processes, collective interactions provide a resource for generating spin-squeezing for metrological applications \cite{defenu2023}, studying nonequilibrium phase transitions \cite{kirton2017,ferioli2023}, and realizing exotic phases of matter \cite{iemini2018,buca2019,russo2025,wang2025}.

\begin{figure}[t]
    \centering
\includegraphics[width=\columnwidth]{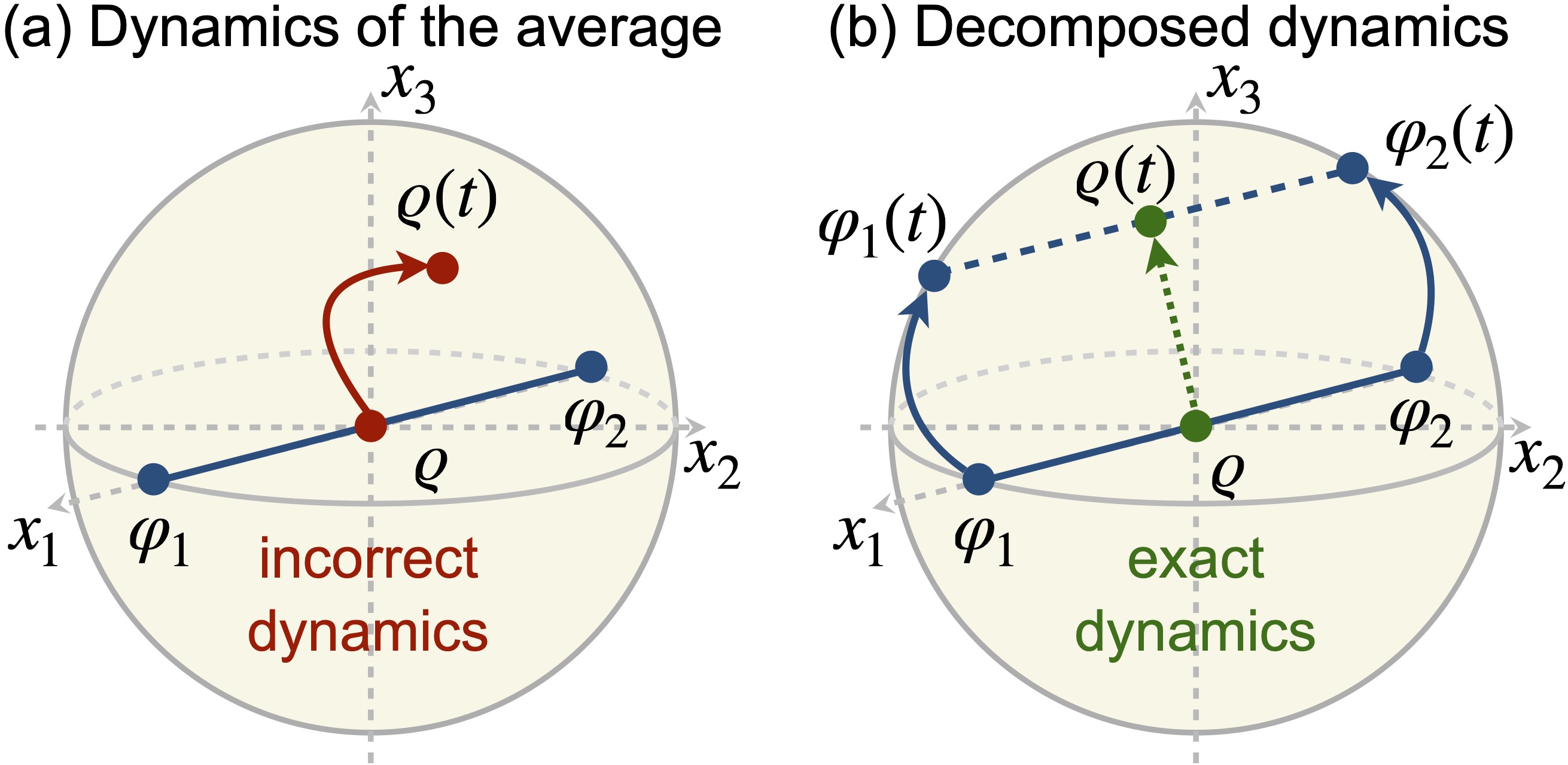}
    \caption{{\bf Decomposed mean-field dynamics.} (a) We consider a strongly correlated state given by a binary macroscopic superposition. At the mean-field level, the two components of the superposition, $\ket{\varphi_1}$ and $\ket{\varphi_2}$, are represented by points on the Bloch sphere. Naively applying mean-field theory defines the average-state density matrix $\varrho$ and directly evolves it through the mean-field equations, typically resulting in an incorrect dynamics. (b) In our framework, the initial superposition state is exactly decomposed into its  components. Each component evolves independently under the mean-field equations (solid arrows), and averaging their evolution yields the exact  dynamics (dashed arrow) at each time.}
\label{fig1}
\end{figure} 

A powerful approach for analyzing these architectures is mean-field theory \cite{spohn1980,defenu2023}. In the {\it strong} long-range regime \cite{defenu2023,mattes2025,winter2026}, each constituent of a many-body ensemble couples collectively to all others. In this scenario, local correlations are suppressed \cite{mattes2025} and mean-field dynamics become exact, reducing the exponential complexity of open many-body quantum dynamics to a small set of nonlinear equations \cite{spohn1980,alicki1983,carollo2021,carollo2024,fiorelli2023}. The exactness is proven assuming initial {\it clustering states}, i.e., states with rapidly decaying correlations \cite{strocchi2020}.
When a long-range interacting system is initialized in a strongly correlated state (e.g., a macroscopic superposition or mixture of distinct phases \cite{winter2026}), the conventional consensus dictates that mean-field theory inherently fails \cite{defenu2023,mukherjee2024,wang2025}, as average expectation values neglect  correlations associated with macroscopically distinct states [see Fig.~\ref{fig1}(a)]. This is compounded by the criticism that mean-field equations do not reflect symmetries of the underlying system. 

As a consequence, approaches to go beyond mean-field theory, such as cumulant expansions (see e.g.~Ref.~\cite{plankensteiner2022}), have been developed. By truncating higher-order correlations at a given order, these methods aim to describe complex phenomena near nonequilibrium phase transitions, where strongly correlated  states emerge. They, however, have several limitations \cite{plankensteiner2022,kerber2025}. For instance, at low orders they approximate strongly correlated states with Gaussian distributions and they only become exact when all cumulants are retained, which is equivalent to solving the full quantum dynamics. Moreover, at finite truncation order, convergence may be non-monotonic and unphysical results can appear \cite{fowler-wright2023,kerber2025}.  
Yet, the necessity of addressing strongly correlated states---emerging in quantum metrology, quantum computing devices, and symmetry-breaking transitions---exposes a fundamental bottleneck in our ability to faithfully simulate large-scale open quantum systems.

In this paper, we make progress in this direction and derive exact results for  the long-range dynamics of strongly correlated states. By exploiting the  convex structure of the space of quantum states in the thermodynamic limit, we show that the evolution of these states can be decomposed into, and entirely reconstructed from, a collection of uncoupled mean-field dynamics [see sketch in Fig.~\ref{fig1}(b)]. This {\it decomposed mean-field approach} naturally accounts for cumulants at all orders. It further allows us to determine, to our knowledge for the first time, a setting in which cumulant expansions are provably exact: for equal-weight binary superpositions or mixtures, they in fact correctly predict second-order cumulants.
To illustrate the relevance of our findings, we analyze two instances of nonequilibrium symmetry breaking. First, we consider symmetry breaking in an open quantum Lipkin-Meshkov-Glick (LMG) model \cite{lipkin1965,morrison2008,lee2014,kothe2026}, describing the exact bimodality of the state and its non-Gaussian fluctuations. Second, we investigate time-translation symmetry breaking in open quantum time crystals \cite{iemini2018,buca2019,carollo2022}, showing  the existence of a {\it nontrivial} stationary state in the time-crystalline phase. These results show that strongly correlated states allow for the restoration of spontaneously broken symmetries in long-range open quantum systems \cite{thirring1968,sheikh2021}. 

Our mean-field decomposition fully reproduces the long-range many-body dynamics, which is out of reach of cumulant expansions at any finite order.  By shifting the perspective from truncation of correlations to decomposition of quantum states, our work redefines the regime of applicability of the mean-field framework. \\

\noindent \textbf{Long-range interacting systems.---} We consider a quantum system made of $N$ components, each consisting of a $d$-level system. The single-component algebra of operators is spanned by a set of Hermitean operators $\{v_\mu\}_{\mu=1}^{d^2}$, chosen such that ${\rm tr}(v_\mu v_\nu)\propto\delta_{\mu\nu}$. We consider here the situation of infinite-range interactions (see below for a discussion of the more general case) and thus introduce the collective operators $V_\mu=\sum_{k=1}^Nv_\mu^{(k)}$. For these systems, the dynamics of operators is governed by the quantum master equation $\dot{X}=\mathbb{L}^*[X]$, with \cite{breuer2002} 
\begin{equation}
\mathbb{L}^*[X]=i[H,X]+\sum_{\mu}\left(J_\mu^\dagger X J_\mu-\frac{1}{2}\{J_\mu^\dagger J_\mu, X\}\right)\, .
\label{Lindblad}
\end{equation}
The Hamiltonian, $H=H_0+H_{\rm int}$, decomposes into a noninteracting term accounting for the free coherent evolution of each single component, $H_0=\sum_{\mu}\omega_\mu  V_\mu$, and an interacting term, $H_{\rm int}=(1/N)\sum_{\mu,\nu}\Omega_{\mu\nu}V_\mu V_\nu$. The jump operators $J_\mu$ can likewise be of two types: collective jump operators of the form $J_\mu=(1/\sqrt{N})\sum_\nu c_\nu V_\nu$, or single-component jump operators $J_{\mu_k}=\sum_{\nu}d_\nu v_\nu^{(k)}$, for each component $k=1,2,\dots N$. The rescalings $1/N,1/\sqrt{N}$ guarantee a well-defined thermodynamic limit \cite{kac1963,benatti2018}. 

When studying such systems, one is often interested in the dynamics of {\it macroscopic observables}, defined as the sample mean $m_\mu^N=V_\mu /N$ \cite{benatti2018}. The action of the generator in Eq.~\eqref{Lindblad} on these operators reads
\begin{equation}
\mathbb{L}^*\left[{m}^N_\mu\right]\approx \sum_{\nu}a_{\mu \nu} m_{\nu}^N +\sum_{\nu\eta} b_{\mu\nu\eta} m_\nu^N m_{\eta}^N \, ; 
\label{Lindblad_application}
\end{equation}
the approximate symbol indicates that there may be additional terms, which are however of order $1/N$ thus becoming irrelevant in the large-$N$ limit \cite{carollo2021,carollo2024}. The equation shows that the dynamics of  the first-moment observables $m_\mu^N$ couples to second-order moments $m_\nu^N m_\eta^N$, giving rise to an infinite hierarchy of coupled equations for higher and higher moments. 

To bypass the hierarchy of equations, a standard approach is to take the quantum expectation of Eq.~\eqref{Lindblad_application} with respect to the time-dependent state $\langle \cdot \rangle_t^\rho:={\rm {Tr}}\left(\rho \, e^{t\mathbb{L}^*}[\cdot]\right)$, and to factorize the quadratic term, i.e., assuming $\langle m_\mu^N m_\nu^N\rangle_t^\rho \mapsto \langle m_\mu^N \rangle_t^\rho\langle m_\nu^N\rangle_t^\rho$. This mean-field approach generates the set of nonlinear equations 
\begin{equation}
\dot{x}_\mu^\rho(t)= \sum_{\nu}a_{\mu \nu} \, x_{\nu}^\rho(t) +\sum_{\nu\eta} b_{\mu\nu\eta} \, x_\nu^\rho(t) x_{\eta}^\rho(t)\, ,
\label{meanfield}
\end{equation}
for the classical variables $x_\mu^\rho(t)$. These are supposed to embody the behavior of the macroscopic observables in the thermodynamic limit. For the models described by the  generator in Eq.~\eqref{Lindblad}, this approach has been shown to be exact when the initial state $\rho$ is {\it clustering} \cite{carollo2021,fiorelli2023,carollo2024}, i.e., with sufficiently weak correlations. Namely, one has   
\begin{equation}
\lim_{N\to\infty}\left\langle \left[m_\mu^N-x_\mu^\rho(t)\right]^2\right\rangle_t^\rho=0\, , \quad \forall t>0, 
\label{clustering}
\end{equation}
whenever the same condition is met at the initial time, $t=0$.  
In the thermodynamic limit, the time-evolved macroscopic operators  therefore  converge to the deterministic variables $x_{\mu}^{\rho}$, at all times, given that the variance in Eq.~\eqref{clustering} vanishes (as in a law of large numbers).

To make this concrete, consider an ensemble of $N$ spin-$1/2$ particles with collective spin operators $S_{\alpha}$, with $\alpha=1,2,3$ and $[S_\alpha,S_\beta]=i\epsilon_{\alpha\beta\gamma}S_\gamma$. Take the state $\rho_1=\ket{\downarrow_N}\!\bra{\downarrow_N}$, with $S_3\ket{\downarrow_N}=-(N/2)\ket{\downarrow_N}$. This state is clustering, with $x_3^{\rho_1}(0)=-1/2$ and all other components being zero. The state  $\rho_2=\ket{\uparrow_N}\!\bra{\uparrow_N}$ behaves analogously with $x_3^{\rho_2}(0)=1/2$. For each of these two states separately, the macroscopic observables can be evolved directly using the mean-field prescription in Eq.~\eqref{meanfield}. 
On the other hand, if we consider the state $\rho=\ket{\psi}\!\bra{\psi}$, with $\ket{\psi}= c_1\ket{\downarrow_N}  + c_2\ket{\uparrow_N}$ ($c_1,c_2\neq0$), the variance of the spin along the third direction [cf.~Eq.~\eqref{clustering}] does not vanish. This means that such a state $\rho=\ket{\psi}\!\bra{\psi}$  violates Eq.~\eqref{clustering} at $t=0$. As a consequence, one is not in principle entitled to use the mean-field equations with initial conditions dictated by $\rho$, and doing so typically leads to an incorrect  dynamical prediction, as discussed in Ref.~\cite{mukherjee2024} [see an illustration in Fig.~\ref{fig1}(a)]. \\

\noindent \textbf{Cumulant expansions.---} This shortcoming has led mean-field theory to be deemed inadequate for such situations, prompting the development of alternative approaches. One method is the so-called {\it cumulant expansion}. The key idea is that mean-field theory can be viewed as setting all second-order cumulants to zero,  precisely leading to the factorization $\langle m_\mu^N m_\nu^N\rangle_t^\rho \mapsto \langle m_\mu^N \rangle_t^\rho\langle  m_\nu^N\rangle_t^\rho$. The cumulant expansion instead retains higher-order cumulants, as follows.  Eq.~\eqref{Lindblad_application} shows that acting on a macroscopic observable produces a polynomial in these operators of first and second order. Acting on a product of two macroscopic observables similarly produces terms up to third order, and so on. The cumulant expansion truncates this hierarchy at a chosen order, by setting all cumulants beyond that order to zero. At the first order beyond mean-field theory, one assumes that third-order cumulants vanish, $K_{\alpha\beta\gamma}^{N}=\langle \!\langle m_\alpha^N m_\beta^N m_\gamma^N\rangle\!\rangle_t\equiv 0$ (the double bra-ket notation denotes cumulants). This allows third-order monomials to be expressed in terms of first- and second-order moments, thereby closing the set of equations. The truncation procedure can in principle be done at any order. 

This approach offers a way to describe  correlations associated with the nature of the initial state. However, there appears to be no systematic control of the approximation error \cite{plankensteiner2022,fowler-wright2023,kerber2025}. While the hope is that retaining higher orders yields an increasingly accurate description, whether adding cumulants actually improves the approximation remains unclear.
In what follows, we shed light on the  actual structure of the dynamics and show how mean-field equations can provide the exact dynamics for the whole hierarchy of cumulants. \\

\noindent \textbf{Decomposed mean-field dynamics.---} Consider the state $\rho=\ket{\Psi}\!\bra{\Psi}$ with $\ket{\Psi}=c_1\ket{\varphi_1}+c_2\ket{\varphi_2}$ and $\ket{\varphi_1},\ket{\varphi_2}$ being clustering states (for example, the states $\ket{\downarrow_N},\ket{\uparrow_N}$ considered above). This state is, in general, not clustering. As such, the mean-field equations \eqref{meanfield}, with initial conditions $x_\mu^\rho(0)=\lim_{N\to\infty}{\rm Tr}(\rho \, m_\mu^N)$, do not correctly provide, in general, the exact dynamics [cf.~Fig.~\ref{fig1}(a)]. Rather than truncating to higher-order cumulants, here we present a modified mean-field approach which is exact in the thermodynamic limit. 

The idea is as follows. For large $N$, the superposition $\ket{\Psi}$ behaves as the statistical mixture $\rho \approx |c_1|^2\ket{\varphi_1}\!\bra{\varphi_1}+|c_2|^2\ket{\varphi_2}\!\bra{\varphi_2}$. The reason is that coherent effects become irrelevant when focussing on the macroscopic observables $m_\mu^N$. Due to the linearity of the dynamical generator in Eq.~\eqref{Lindblad}, we write 
\begin{equation}
e^{t\mathbb{L}}[\rho]\approx |c_1|^2 e^{t\mathbb{L}}\left[\ket{\varphi_1}\!\bra{\varphi_1}\right]+|c_2|^2e^{t\mathbb{L}}\left[\ket{\varphi_2}\!\bra{\varphi_2}\right]\, ,
\label{linearity}
\end{equation}
where $\mathbb{L}$ is the generator in the Schrödinger picture. 
This relation suggests that the exact evolution of the system can be obtained by calculating the mean-field dynamics for each state $\rho_k=\ket{\varphi_k}\!\bra{\varphi_k}$, and combining them through the probabilities $|c_k|^2$. As an example, one is tempted to write 
$$
\lim_{N\to\infty}{\rm Tr}\left(e^{t\mathbb{L}}[\rho] \, m_\mu^N\right)= |c_1|^2 x_\mu^{\rho_1}(t)+|c_2|^2 x_\mu^{\rho_2}(t)\, , 
$$
with $x_\mu^{\rho_k}(t)$ being the solution of the mean-field dynamics in Eq.~\eqref{meanfield}, with initial state $\rho_k$. As summarized in the following theorem, this procedure is 
exact and provides all moments and cumulants of the macroscopic observables.  

\begin{theorem-non}[Decomposed  dynamics] Let us consider the dynamics induced by Eq.~\eqref{Lindblad} and focus on the time evolution of the macroscopic observables $m^N_\mu$. Then, the following statements hold:

$i)$ The superposition $\ket{\Psi}=\sum_{k}c_k \ket{\varphi_k}$, where the states $\ket{\varphi_k}$ are clustering and such that $\lim_{N\to\infty}\braket{ \varphi_k|\varphi_h}=0$, is indistinguishable from the statistical mixture 
$$
\ket{\Psi}\!\bra{\Psi}\, \leftrightarrow\, \rho=\sum_{k}|c_k|^2 \ket{\varphi_k}\!\bra{\varphi_k}\, ;
$$

$ii)$ With $\rho=\sum_k P_k \rho_k$, where $P_k\ge0$, $\sum_k P_k=1$, and $\rho_k$ being pure or mixed clustering states, the time-evolved moment generating function of the macroscopic observables can be computed as  
\begin{equation}
        \lim_{N\to\infty}{\rm Tr}\left(\rho \, e^{t\mathbb{L}^*}[e^{-\sum_\mu s_\mu m_\mu^N}]\right)\!=\! \sum_{k}P_k e^{-\sum_\mu s_\mu x_\mu^{\rho_k}(t)}\,  ,
        \label{decomposed_MGF}
    \end{equation}
where $x_\mu^{\rho_k}(t)$ is the  solution of Eq.~\eqref{meanfield} with initial value computed from $\rho_k$. An analogous relation holds for regular functions the macroscopic observables. 
\end{theorem-non}

This result (the proof is reported in the Supplemental Material \cite{SM}) shows that the complete statistical characterization of the macroscopic observables, for strongly correlated initial states, is obtained from the mean-field equations. A direct consequence of the above theorem, which exactly captures the structure of the long-range  quantum dynamics, is the possibility to derive a first rigorous statement on the applicability of cumulant expansions. As detailed below, for equal-weight binary superpositions or statistical mixtures, a second-order cumulant expansion correctly describes second-order cumulants.

\begin{figure*}[t]
    \centering
\includegraphics[width=\textwidth]{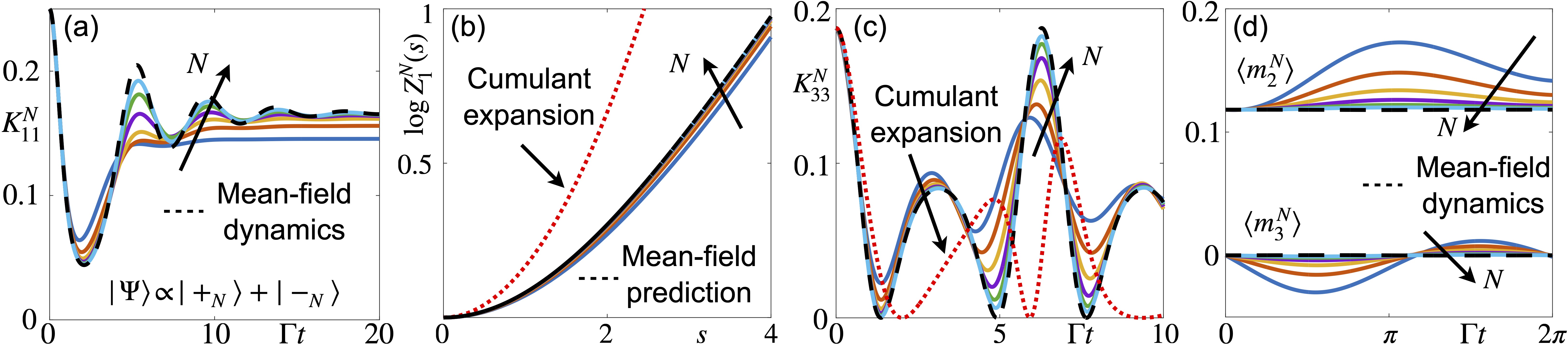}
    \caption{{\bf Nonequilibrium systems with symmetry breaking.} (a) Evolution of the second cumulant $K_{11}^N$ (variance of $m_1^N$) for the open LMG model, starting from $\ket{\Psi}\propto \ket{+_N}+\ket{-_N}$. The dashed line is the decomposed mean-field dynamics, while the solid lines are numerical results. We set $\omega/\Gamma=1$ and $\Omega/\Gamma=2$, for which the model is in the symmetry-broken phase. (b) Stationary value of the full cumulant generating function of $m_1^N$ for the model in (a). We compare it with the Gaussian prediction of the second-order cumulant expansion (dotted line). (c) Evolution of the second cumulant $K_{33}^N$ (variance of $m_3^N$) for the boundary time-crystal, starting from $\ket{\Psi}\propto \ket{\uparrow_N}+\sqrt{3}\ket{\downarrow_N}$. The dashed line is the decomposed mean-field dynamics, while the dotted one is the prediction from the second-order cumulant expansion. Here, $\omega/\Gamma=\sqrt{5/4}$. (d) Benchmark of the time-invariance of the state in Eq.~\eqref{t-inv-btc} for the model in (c). In all panels, numerical results are for $N=16,32,64,128,256,512$. }
\label{fig2}
\end{figure*}

\begin{corollary-non}
Consider a binary  superposition $\rho=\ket{\Psi}\!\bra{\Psi}$, with $\ket{\Psi}=c_1\ket{\varphi_1}+c_2\ket{\varphi_2}$, such that $|c_1|^2=|c_2|^2=1/2$, or a statistical mixture $\rho=P_1\rho_1+P_2\rho_2$, such that $P_1=P_2=1/2$, with  $\ket{\varphi_k}$ and $\rho_k$ being clustering. Under the dynamics generated by Eq.~\eqref{Lindblad}, we have ($\forall t\ge0$)
$$
K_{\alpha\beta\gamma}=\lim_{N\to\infty}\left\langle\!\left\langle m_\alpha^N m_\beta^N m_\gamma^N\right\rangle\!\right\rangle^{\rho}_t =0\,, \quad \forall \alpha,\beta,\gamma \, . 
$$
\end{corollary-non}
A derivation of the above result is given in Ref.~\cite{SM}. While it provides a regime in which second-order cumulant expansions are exact, it also implies their redundancy. To prove it, we used the decomposed mean-field approach, which is numerically more efficient, is more amenable to analytical solutions and  provides the full hierarchy of moments and cumulants  [cf.~Eq.~\eqref{decomposed_MGF}]. \\

\noindent \textbf{Application 1: open LMG model.---} We consider an open quantum LMG model that displays a nonequilibrium symmetry-breaking phase transition \cite{SM}. The Hamiltonian is $H=\omega S_3 - (\Omega/N) S_1 S_1 $ and the jump operator is $J=\sqrt{\Gamma/N}S_-$, with $S_-=(S_1-iS_2)$. The system possesses a weak $\mathcal{Z}_2$ symmetry, with parity operator $P=e^{i\pi S_3}$ ($S_{1/2}\to -S_{1/2}$) and the mean-field equations of the model (given in Ref.~\cite{SM}) reflect such symmetry. What typically breaks the symmetry in the mean-field evolution is the choice of a clustering initial state. Our findings, however, permit to consider dynamics starting from symmetric (strongly correlated) states, effectively  restoring the symmetry of the mean-field dynamics. 

As an example, we take the state $\rho=\ket{\Psi}\!\bra{\Psi}$, with
$\ket{\Psi}\propto \ket{+_N}+\ket{-_N}$
and $S_1\ket{\pm_N}=\pm (N/2)\ket{\pm_N}$.  We analyze the evolution of the second-order cumulant $K_{11}^N=\langle\!\langle (m_1^N)^2\rangle\!\rangle_t^{\rho}$ in the symmetry-broken phase, comparing it with the decomposed mean-field dynamics [cf.~Fig.~\ref{fig2}(a)]. Our corollary implies that the second-order cumulant expansion also correctly captures the variance for such a state. However, in Fig.~\ref{fig2}(b) we show that this approach is off when predicting the stationary cumulant generating function of $m_1^N$, $\log Z_1^N(s)$ with $Z_1^N(s)=\langle e^{-s\, m_1^N}\rangle^\rho_t$. It indeed provides a Gaussian approximation, neglecting the bimodality of the stationary state stemming from the symmetry breaking. \\

\noindent \textbf{Application 2: boundary time-crystal.---} We consider the boundary time-crystal \cite{iemini2018} which displays a nonequilibrium phase transition from a stationary phase to time-crystal limit-cycle dynamics \cite{iemini2018,buca2019}. The  Hamiltonian is $H=\omega S_1$ and the jump operator is as before. The mean-field equations are given in Ref.~\cite{SM}. We assume that the initial state is such that $\langle m_1^N\rangle=0$. For $\omega/\Gamma\le 1/2$, the equations admit a stable stationary solution \cite{iemini2018,carollo2022}. On the other hand, for $\omega/\Gamma>1/2$ there is no admissible stationary state and the equations give rise to limit-cycle dynamics \cite{iemini2018}. In this regime, one has (with $\rho_{\tau_0}=\ket{\varphi_{\tau_0}}\!\bra{\varphi_{\tau_0}}$ being a pure clustering state) 
\begin{equation}
    x_2^{\rho_{\tau_0}}(t)=\frac{1}{2}\cos \theta(t+\tau_0)\, ,\quad  x_3^{\rho_{\tau_0}}(t)=\frac{1}{2}\sin \theta(t+\tau_0)\, . 
\end{equation}
Here, $\theta(t+\tau_0)$ (given in Ref.~\cite{SM}) is a highly nonlinear function of $t+\tau_0$ with period $T=2\pi/D$ and $D=\sqrt{(\omega/\Gamma)^2-1/4}$. The parameter $\tau_0$ is such that $\theta(\tau_0)$ gives the initial condition associated with  $\rho_{\tau_0}$. 

In Fig.~\ref{fig2}(c), we show the dynamics of the variance $K_{33}^N$ for the initial state $\ket{\Psi}=(\ket{\uparrow_N}+\sqrt{3}\ket{\downarrow_N})/2$. We observe that our decomposed mean-field dynamics describes the evolution in the thermodynamic limit, while the second-order cumulant expansion \cite{SM} is not  accurate. Exploiting our analytical results, we furthermore show how to construct a stationary state within the time-crystal regime ($\omega/\Gamma>1/2$). We define the state 
\begin{equation}
\rho=\frac{1}{{T}}\int_0^{T}{\rm d} \tau_0 \, \rho_{\tau_0} , 
\label{t-inv-btc}
\end{equation}
providing a state averaged over a cycle of the asymptotic dynamics. This construction resembles symmetry restoration for equilibrium mean-field models \cite{thirring1968,sheikh2021}; here restoring the time-translation symmetry requires integrating the initial condition over a cycle, in analogy to $U(1)$ symmetries \cite{thirring1968}. 
Using the state in Eq.~\eqref{t-inv-btc}, we compute the expectation of a generic function $F=F[m_2^N,m_3^N]$, at any time $t$. We find
$$
\lim_{N\to\infty}{\rm Tr}\left[\rho  F\right]=\frac{1}{T}\int_0^{T}\!\!{\rm d}\tau_0 \, F[x_2^{\rho_{\tau_0}}(t),x_3^{\rho_{\tau_0}}(t)]\, .
$$
Since the integrand is periodic in $t+\tau_0$, integrating $\tau_0$ over a  period makes the expectation independent on $t$. The state $\rho$ thus behaves as a time-translation symmetric quasi-stationary state \cite{kastner2011,defenu2021}. We  demonstrate this numerically for two  expectation values in Fig.~\ref{fig2}(d). Analogous results hold when considering $\ket{\Psi}\propto \int_0^T {\rm d}\tau_0 \ket{\varphi_{\tau_0}}$. In that case, however, convergence with $N$ is slower \cite{SM}. 
\\

\noindent \textbf{Discussion.---} We have presented an exact approach that efficiently provides the dynamics of long-range interacting systems starting from strongly correlated states. Our approach entirely  bypasses the truncation framework of the most widely employed cumulant expansions,  which, as we have also discussed, suffer from several limitations \cite{plankensteiner2022,fowler-wright2023,kerber2025}. We have focused on systems with infinite-range interactions; however our findings extend to generic many-body systems in the strong long-range regime. That is, when components are coupled with an interaction strength decaying as an inverse power of the distance, with exponent smaller than or equal the dimensionality of the system \cite{defenu2023,mattes2025,winter2026}. In analogy to what presented here, in those cases the time-evolved reduced density matrix for local observables \cite{spohn1980,alicki1983,mattes2025}, starting from a strongly correlated state,   coincides with the weighted sum of the evolution of the reduced density matrices obtained from the individual clustering states [cf.~Fig.~\ref{fig1}(b)]. 

While our derivation shows that cumulant expansions are unnecessary for long-range interacting systems in the strong long-range regime,  these approximate approaches---together with other approximation methods \cite{hosseinabadi2025,noel2026}---remain relevant for {\it shorter-ranged} interactions. There, mean-field theory fails and large-scale numerical methods can struggle, particularly in dimensions higher than one, so that these approaches provide a viable tool to explore many-body dynamics.  
\\

\textbf{Data availability.---} The data displayed in this manuscript is
available on Zenodo \cite{zenodo}.

\bibliography{refs.bib}

\newpage
\setcounter{equation}{0}
\setcounter{figure}{0}
\setcounter{table}{0}
\makeatletter
\renewcommand{\theequation}{S\arabic{equation}}
\renewcommand{\thefigure}{S\arabic{figure}}
\makeatletter

\onecolumngrid
\newpage

\setcounter{page}{1}
\begin{center}
{\Large SUPPLEMENTAL MATERIAL}
\end{center}
\begin{center}
\vspace{0.8cm}
{\Large Don't truncate, decompose: mean-field dynamics of long-range quantum systems from strongly correlated states}
\end{center}
 \begin{center}
 Federico Carollo$^1$
 \end{center}
 \begin{center}
 $^1${\em Dipartimento di Fisica, Sapienza Università di Roma, Piazzale Aldo Moro 5, 00185 Rome, Italy}\\
 \end{center}

 \author{Federico Carollo}
\affiliation{Dipartimento di Fisica, Sapienza Università di Roma, Piazzale Aldo Moro 5, 00185 Rome, Italy}

\section{I. Proof of theorem}
Let us consider the macroscopic superposition $\ket{\Psi}$ presented in the main text and use it to compute the moment generating function of the macroscopic observables $m^N=(m_1^N,m_2^N,\dots m^{d^2}_N)$. We have 
\begin{equation}
Z(s)=\lim_{N\to\infty}\sum_{k,h}c_h^* c_k\bra{\varphi_h}e^{t\mathbb{L}^*}\left[e^{-\sum_{\mu}s_\mu m_\mu^N}\right]\ket{\varphi_k} \, .
\label{pure-mixed}
\end{equation}
The double summation is bounded for any $N$, so that we can bring the large-$N$ limit in front of the expectation value. To prove point $i)$ of the theorem, we just need to show that the off-diagonal elements of the double sum vanish in the thermodynamic limit. We construct the following quantity 
$$
I_{hk}=\bra{\varphi_h}\left(e^{t\mathbb{L}^*}\left[e^{-\sum_{\mu}s_\mu m_\mu^N}\right]-e^{-\sum_{\mu}s_\mu x_\mu^{\varphi_k}(t)}\right)\ket{\varphi_k}\, ,
$$
where $x_\mu^{\varphi_k}(t)$ is the solution of Eq.~\eqref{meanfield} with initial conditions given by the clustering pure state $\varphi_k=\ket{\varphi_k}\!\bra{\varphi_k}$. The difference contained in the round brackets in the equation above can be written as
\begin{equation}
e^{-\sum_\mu s_\mu m_\mu^N}-e^{-\sum_{\mu}s_\mu x_\mu^{\varphi_k}(t)}=-\sum_\nu s_\nu \int_0^{1}{\rm d } u \, e^{-u\sum_{\mu} s_\mu m_\mu^N }e^{-(1-u)\sum_\mu s_\mu x_\mu^{\varphi_k}(t)}\left(m_\nu^N-x_\nu^{\varphi_k} (t)\right)\, .
    \label{exp-exp}
\end{equation}
We apply the evolution $e^{t\mathbb{L}^*}$ on the expression above and take the matrix element using the states $\ket{\varphi_h},\ket{\varphi_k}$. By taking the Cauchy-Schwarz inequality, we find
$$
|I_{hk}|\le d^{2} s_{\rm max} e^{d^2 s_{\rm max} M}\max_{\forall \nu}\left[\sqrt{\bra{\varphi_k} e^{t\mathbb{L}^*}\left[\left(m_\nu^N-x_\nu^{\varphi_k}(t)\right)^2\right]\ket{\varphi_k}}\right]\, .
$$
Here, $s_{\rm max}:=\max_{\forall \mu}\{|s_\mu|\}$ and $M:=\max_{\forall\mu}\{ \|v_\mu\|\}$. Note that due to Lemma 4 in Ref.~\cite{fiorelli2023}, we have that $|x_\mu^{\varphi_k}(t)|\le \|v_\mu\|\le M$.  
Since the state $\varphi_k$ is clustering and $x_\nu^{\varphi_k}(t)$ is the solution of the corresponding mean-field equations, we have that $\lim_{N\to\infty}|I_{kh}|=0$ \cite{carollo2021,carollo2024,fiorelli2023} (see also discussion below in this proof). This means that 
$$
\lim_{N\to\infty}\bra{\varphi_h}e^{t\mathbb{L}^*}\left[e^{-\sum_{\mu}s_\mu m_\mu^N}\right]\ket{\varphi_k}=e^{-\sum_{\mu}s_\mu x_\mu^{\varphi_k}(t)}\lim_{N\to\infty} \braket{\varphi_h|\varphi_k}=0\, ,
$$
implying that all off-diagonal terms in the double sum of Eq.~\eqref{pure-mixed} vanish and the quantum superposition state is indistinguishable from the statistical mixture of the single components. \\

We now move to the proof of the second point of the theorem, for which we will use some of the ideas already used for point $i)$.
Let us consider a quantum state for a finite-$N$ system which is given by the statistical mixture 
\begin{equation}
    \rho=\sum_k P_k \rho_k\, , 
\end{equation}
with $\sum_k P_k=1$ and with $\rho_k$ being pure and/or mixed clustering states, in the thermodynamic limit. That is, we have (at the initial time) that 
$$
\lim_{N\to\infty} \left\langle \left(m_\mu^N-x_\mu^{\rho_k}\right)^2\right\rangle^{\rho_k}=0\, , \qquad \mbox{ where } \qquad x_\mu^{\rho_k}=\lim_{N\to\infty}\langle m_\mu^N\rangle^{\rho_k}\, ,
$$
with the definition $\langle O\rangle^{\rho_k}:={\rm Tr}\left(\rho_k O\right)$. 

To prove point $ii)$, we consider the time evolved expectation value 
\begin{equation}
\left\langle e^{t\mathbb{L}^*}\left[e^{-\sum_\mu s_\mu m_\mu^N}\right]\right\rangle^\rho=\sum_{k}P_k \left\langle e^{t\mathbb{L}^*}\left[e^{-\sum_\mu s_\mu m_\mu^N}\right]\right\rangle^{\rho_k}=\sum_{k}P_k \left\langle e^{-\sum_\mu s_\mu m_\mu^N}\right\rangle^{\rho_k}_t\, ,
\end{equation}
where, for the sake of compactness, we introduced the notation 
$$
\left\langle e^{t\mathbb{L}^*}\left[O\right]\right\rangle^{\rho_k}=\left\langle O\right\rangle^{\rho_k}_t\, .  
$$
We need to prove that 
\begin{equation}
\begin{split}
\lim_{N\to\infty}\sum_k P_k\left\{\left\langle e^{-\sum_\mu s_\mu m_\mu^N}\right\rangle^{\rho_k}_t-e^{-\sum_\mu s_\mu x_\mu^{\rho_k}(t)}\right\}=0 \, .
\end{split}
\label{SM:thesis_1}
\end{equation}
Since the summation is bounded for any $N$, we can bring the limit inside the sum over $k$. We can therefore control the behavior of each term separately.  Using the relation in Eq.~\eqref{exp-exp}, we write 
$$
I_k^N:=\left\langle e^{-\sum_\mu s_\mu m_\mu^N}\right\rangle^{\rho_k}_t-e^{-\sum_\mu s_\mu x_\mu^{\rho_k}(t)} = -\sum_\nu s_\nu \int_0^1 {\rm d}u \, \left\langle e^{-u\sum_{\mu} s_\mu m_\mu^N }e^{-(1-u)\sum_\mu s_\mu x_\mu^{\rho_k}(t)}\left(m_\nu^N-x_\nu^{\rho_k} (t)\right)\right\rangle^{\rho_k}_t \, .
$$
Taking the modulus of $I_k^N$ and applying the Cauchy-Schwarz inequality, we find the bound
\begin{equation}
\left|I_k^N\right|\le d^2 s_{\rm max }\, e^{d^{2} s_{\rm max} M } \max_{\forall \nu}\left[\sqrt{\left\langle \left[m_\nu^N - x_\nu^{\rho_k}(t)\right]^2\right\rangle^{\rho_k}_t}\right]\, .
\label{SM:bound}
\end{equation}
Here, $s_{\rm max}:=\max_{\forall \mu}\{|s_\mu|\}$ and $M:=\max_{\forall \mu}\{ \|v_\mu\|\}$. Note that due to Lemma 4 in Ref.~\cite{fiorelli2023}, we have that $|x_\mu^{\rho_k}(t)|\le \|v_\mu\|\le M$.  
Since the states $\rho_k$ are clustering and since the variables $x_\nu^{\rho_k}(t)$ are the solutions of the mean-field equations with initial conditions given by the state $\rho_k$ itself, we have that $\lim_{N\to\infty}I_k^{N}=0$ (see, e.g., main theorems in Refs.~\cite{carollo2021,carollo2024,fiorelli2023}). This, in turn, implies the validity of the relation in Eq.~\eqref{SM:thesis_1}. \\

In the main text, we also mention that a similar relation holds for generic functions of the macroscopic observables. We explain here the idea. Consider a generic operator-valued function $F$ of the vector $m^N=(m_1^N,m_2^N,\dots,m_{d^2}^N)$ collecting all macroscopic observables. We assume that the function admits a multivariate Taylor expansion so that it can be written as 
$$
F(m^N)=\sum_{\ell_1,\ell_2,\dots \ell_{d^2}} C_{\ell_1,\ell_2,\dots \ell_{d^2}} \left(m_1^{N}\right)^{\ell_1}\left(m_2^{N}\right)^{\ell_2}\dots \left(m_{d^2}^{N}\right)^{\ell_{d^2}}\, .
$$
We assume that the series converges within a domain $\mathcal{D}=[-M,M]^{d^2}$. (This is not strictly necessary as we could also consider functions only defined up to a given time; we make this assumption to simplify the treatment.)
We take the expectation value of the function and consider the large-$N$ limit. Given that the summation over the states $\rho_k$  is bounded for any $N$, we find 
$$
\lim_{N\to\infty}\langle F(m^N) \rangle_t^\rho = \sum_{k} P_k \lim_{N\to\infty}\langle F(m^N)\rangle_t^{\rho_k}\, . 
$$
We then compare each expectation value taken with respect to the clustering states $\rho_k$ against the function $F$ evaluated using the vector $x^{\rho_k}$ of the mean-field variables. We have 
\begin{equation}
\begin{split}
D_k^N&:=\langle F(m^N)\rangle_t^{\rho_k}-F(x^{\rho_k}(t))\\
&=\!\!\sum_{\ell_1,\ell_2,\dots \ell_{d^2}} \!\!\!C_{\ell_1,\ell_2,\dots \ell_{d^2}} \left\{\left\langle \left(m_1^{N}\right)^{\ell_1}\left(m_2^{N}\right)^{\ell_2}\dots \left(m_{d^2}^{N}\right)^{\ell_{d^2}}-\left(x_1^{\rho_k}(t)\right)^{\ell_1}\left(x_2^{\rho_k}(t)\right)^{\ell_2}\dots \left(x_{d^2}^{\rho_k}(t)\right)^{\ell_{d^2}}\right\rangle^{\rho_k}_t\right\}\, ,
\end{split}
\end{equation}
which needs to be considered in the thermodynamic limit. Since we have assumed that the Taylor series converges within the maximum values allowed for $m^N$ and $x^{\rho_k}$, we can bring the large-$N$ limit in front of the curly brackets. Then, one can proceed by substituting one at a time the elements of $m^N$ with the elements of $x^{\rho_k}(t)$ starting, for instance, from the left and applying to each substitution the Cauchy-Schwarz inequality to get a bound similar to the one in Eq.~\eqref{SM:bound}. For the sake of concreteness, we show here one example of a such a procedure for a monomial of degree $3$. Let us focus on the difference 
$$
T^N=\langle m_\alpha^N m_\beta^N m_\gamma^N-x_\alpha^{\rho_k}(t)x_\beta^{\rho_k}(t)x_\gamma^{\rho_k}(t)\rangle_t^{\rho_k}\, .
$$
We start by substituting $m_\alpha^N\leftrightarrow x_{\alpha}^{\rho_k}(t)$ to get
$$
T^N=\left\langle \left(m_\alpha^N -x_\alpha^{\rho_k}(t)\right)m_\beta^N m_\gamma^N\right\rangle^{\rho_k}_t+x_\alpha^{\rho_k}(t)\left\langle m_\beta^N m_\gamma^N -x_\beta^{\rho_k}(t)x_\gamma^{\rho_k}(t)\right\rangle_t^{\rho_k}\, .
$$
The first term is already in the proper shape and we see that the second one is now essentially the difference between monomials of lower degree than the original one. We therefore proceed also for this term by substituting $m_\beta^N\leftrightarrow x_{\beta}^{\rho_k}(t)$ and $m_\gamma^N\leftrightarrow x_{\gamma}^{\rho_k}(t)$. This gives 
$$
T^N=\left\langle \left(m_\alpha^N -x_\alpha^{\rho_k}(t)\right)m_\beta^N m_\gamma^N\right\rangle^{\rho_k}_t+x_\alpha^{\rho_k}(t)\left\langle\left(m_\beta^N-x_\beta^{\rho_k}(t)\right)m_\gamma^N \right\rangle_t^{\rho_k} + x_\alpha^{\rho_k}(t)x_\beta^{\rho_k}(t)\left\langle m_\gamma^N-x_\gamma^{\rho_k}(t)\right \rangle^{\rho_k}_t \, ,
$$
which is bounded by 
$$
|T^N|\le M^2 \sqrt{\left\langle \left(m_\alpha^N -x_\alpha^{\rho_k}(t)\right)^2\right\rangle_t^{\rho_k}}+M^2 \sqrt{\left\langle \left(m_\beta^N -x_\beta^{\rho_k}(t)\right)^2\right\rangle_t^{\rho_k}}+M^2 \sqrt{\left\langle \left(m_\gamma^N -x_\gamma^{\rho_k}(t)\right)^2\right\rangle_t^{\rho_k}}\, . 
$$
Such a term vanishes in the large-$N$ limit as discussed  below Eq.~\eqref{SM:bound}.

\section{II. Exactness of second-order cumulant expansion for equal-weight binary statistical mixtures or macroscopic superpositions}
In this section, we provide a proof of the corollary presented in the main text. In particular, we compute a generic third-order cumulant for binary macroscopic superpositions or statistical mixtures. We then show that this is identically zero, in the thermodynamic limit, when the weights of the superposition/mixture are equal. 

Let us consider a state which is a statistical mixture of two clustering states, namely 
\begin{equation}
    \rho=p \rho_1 +(1-p)\rho_2\, , \quad \mbox{ with } \quad  0\le p\le 1 \, .
    \label{SM:mixture}
\end{equation}
We can compute the expectation values of the macroscopic operators for the two states $\rho_1,\rho_2$, in the thermodynamic limit, as 
$$
x_\alpha^{\rho_{1/2}}=\lim_{N\to\infty}{\rm Tr}\left(m_\alpha^N \rho_{1/2}\right)\, .
$$
Given the binary mixture in Eq.~\eqref{SM:mixture}, the first three moments of the macroscopic operators read 
\begin{equation}
\begin{split}
    &\lim_{N\to\infty}\langle m_\alpha^N \rangle =  p x_\alpha^{\rho_1}+(1-p)x_\alpha^{\rho_2}\, , \quad \lim_{N\to\infty}\langle m_\alpha^N m_\beta^N \rangle =  p x_\alpha^{\rho_1} x_\beta^{\rho_1}+(1-p)x_\alpha^{\rho_2} x_\beta^{\rho_2} \, ,\\
    &\lim_{N\to\infty}\langle m_\alpha^N m_\beta^N m_\gamma^N\rangle=p x_\alpha^{\rho_1} x_\beta^{\rho_1} x_\gamma^{\rho_1} +(1-p)x_\alpha^{\rho_2} x_\beta^{\rho_2} x_\gamma^{\rho_2}\, .
    \end{split}
    \label{SM:moments}
\end{equation}
With the above moments, we now compute an expression for a generic  third-order cumulant. In the thermodynamic limit, such a cumulant $K_{\alpha \beta \gamma}$ is given by 
\begin{equation}
K_{\alpha\beta\gamma}=\lim_{N\to\infty}\left(\langle m_\alpha^N m_\beta^N m_\gamma^N\rangle-\langle m_\alpha^N m_\beta^N \rangle\langle m_\gamma^N\rangle-\langle m_\alpha^N m_\gamma^N \rangle\langle m_\beta^N\rangle-\langle m_\beta^N m_\gamma ^N \rangle\langle m_\alpha^N\rangle+2\langle m_\alpha^N\rangle \langle m_\beta^N \rangle \langle m_\gamma^N\rangle\right)\, .
\end{equation}
Substituting the expressions in Eq.~\eqref{SM:moments} into the above equation and collecting  terms, we find
\begin{equation}
\begin{split}
K_{\alpha\beta\gamma}&=\left(p-3p^2 +2p^3\right)x_\alpha^{\rho_1}x_\beta^{\rho_1}x_\gamma^{\rho_1}+\left[-p(1-p)+2p^2(1-p)\right]\left(x_\alpha^{\rho_1}x_\beta^{\rho_1}x_\gamma^{\rho_2}+x_\alpha^{\rho_1}x_\beta^{\rho_2}x_\gamma^{\rho_1}+x_\alpha^{\rho_2}x_\beta^{\rho_1}x_\gamma^{\rho_1}\right)\\
&\!+\!\left[-p(1-p)+2(1-p)^2p\right]\left(x_\alpha^{\rho_2}x_\beta^{\rho_2}x_\gamma^{\rho_1}+x_\alpha^{\rho_2}x_\beta^{\rho_1}x_\gamma^{\rho_2}+x_\alpha^{\rho_1}x_\beta^{\rho_2}x_\gamma^{\rho_2}\right)+\left[(1-p)-3(1-p)^2+2(1-p)^3\right]x_\alpha^{\rho_2}x_\beta^{\rho_2}x_\gamma^{\rho_2} .
\end{split}
\end{equation}
One can check that for $p=1/2$ the generic third-order cumulant is identically zero. Moreover, since the time evolution preserves the weights of the statistical mixture and solely changes the values of $x_\alpha^{\rho_{1/2}}$ for each branch, third-order cumulants remain zero throughout the time evolution.

Finally, we note that since binary macroscopic superpositions of (asymptotically orthogonal) clustering states behave as the statistical mixtures considered in this section, third-order cumulants are identically zero also when the initial state consists of a binary superposition of states with weights having equal modulus. 

\section*{III. Details on the considered models}
In this section, we provide details on the equations for the considered models and the associated symmetry breaking. \\

\noindent {\bf Open quantum LMG model.} The mean-field equations for the LMG model are 
\begin{equation}
\begin{split}
    \dot{x}_1(t)&=-\omega\,  x_2(t)+\Gamma \,x_1(t) x_3(t)\, ,\\
    \dot{x}_2(t)&=\omega\, x_1(t) +2 \,\Omega x_1(t) x_3(t)+\Gamma \,x_2(t) x_3(t)\, ,\\
    \dot{x}_3(t)&=-2\Omega \,x_1(t) x_2(t) -\Gamma \left[x_1^2(t)+x_2^2(t)\right]\, .
    \end{split}
    \label{mfLMG}
\end{equation}
To simplify the discussion, we introduce dimensionless parameters obtained by redefining $\omega \mapsto \omega /\Gamma $, $\Omega\mapsto  \Omega/\Gamma$, $\Gamma \mapsto 1$, and $\Gamma t\mapsto t$. Furthermore, we consider $\omega =1$ and $\Omega >0$. 
We want to compute the stationary solutions of the above equations. Firstly, we note that $x=(0,0,\pm 1/2)$ are fixed points. 

We then check for the existence of solutions with $x_{1/2}\neq0$. With the above parameters and setting Eq.~\eqref{mfLMG} to zero, we  see that $x_3=x_2/x_1$. Looking at the third equation, setting it to zero, and dividing it by $x_1^2$, we find 
$$
x_3^2+2\Omega x_3+1 = 0\, .
$$
Solving for $x_3$, we find two possible solutions $x_3=-\Omega \pm \sqrt{\Omega^2-1}$. Note that, these are real only for $\Omega\ge 1$ and the only one that can match the bounds on $x_3$ is $x_3=-\Omega+\sqrt{\Omega^2-1}$. The latter becomes physical when $x_3\ge -1/2$, which occurs at $\Omega = 5/4$. From the constraint $x_1^2+x_2^2+x_3^2=1/4$ that holds for collective models we can write
$$
x_1^2+x_2^2+x_3^2=x_1^2+x_1^2 x_3^2+x_3^2 =\frac{1}{4}\, ,
$$
which we can solve to find 
$$
x_1=\pm \sqrt{\frac{\frac{1}{4}-x_3^2}{1+x_3^2}}\, .
$$
Taken together, this shows that for $\Omega\ge 5/4$, there appears a solution with $x_3\ge - 1/2$ and two possible values $x_1$. These solutions break the symmetry of the dynamical generator. \\

\noindent {\bf Boundary time-crystal.} We consider similarly as above dimensionless parameters (in units of $\Gamma$) and focus on the situation where $x_1(0)\equiv 0$, which is a conserved quantity \cite{iemini2018,carollo2022}. Without loss of generality we consider $\omega>0$. The mean-field equations for $x_2(t)$ and $x_3(t)$ read 
\begin{equation}
\begin{split}
    \dot{x}_2(t)&=-[\omega-x_2(t)] x_3(t) \, \\
    \dot{x}_3(t)&=[\omega -x_2(t)]x_2(t)\, .
    \end{split}
    \label{mfBTC}
\end{equation}
To find the stationary solution of the model, we set  $x_2=\omega$ and exploit the constraint $x_2^2+x_3^2=1/4$ to recover $x_3$. This gives $x_2=\omega$ and $x_3=\pm \sqrt{1/4-\omega^2}$ which solely exist for $\omega\le 1/2$, with only the negative solution being stable. For $\omega>1/2$, there is no stationary solution and the mean-field equations display asymptotic limit-cycle dynamics. To describe the oscillatory solutions, we define 
$$
x_2(t)=\frac{1}{2}\cos \theta(t+\tau_0)\, , \quad x_3(t)=\frac{1}{2} \sin \theta(t+\tau_0)\, ,\qquad \mbox{ with } \qquad \dot{\theta}=\omega -\frac{1}{2} \cos \theta\, .
$$
The above differential equation can be solved and one has 
$$
\theta(t+\tau_0)=2\tan^{-1}\left[\frac{1}{C(\omega)}\tan \left(\frac{D(\omega) (t+\tau_0)}{2} \right)\right]\, , \, \mbox{ with } \,  C(\omega)=\sqrt{\frac{\omega+\frac{1}{2}}{\omega-\frac{1}{2}}} \, \mbox{ and }\, D(\omega)=\sqrt{\omega^2-\frac{1}{4}}\, .
$$ 
Here, $\tau_0$ is a parameter which is chosen in such a way that $\theta(\tau_0)$ encodes the initial condition. The above solution is an  oscillatory one with period $T=2\pi/D(\omega)$.

\section*{IV. Calculation of cumulant-expansion equations}
In this section, we derive the equations of motion governing the dynamics of the boundary time-crystal under the assumptions that third-order cumulants of the macroscopic observables are zero. We first calculate terms associated with the Hamiltonian and then those associated with the dissipative part of the dynamical generator. The computation requires working out the action of the generator on one-point and two-point functions of the observables. 

When looking at the one-point operators, the Hamiltonian contribution gives rise to the relations
\begin{equation}
\begin{split}
&i[H,S_1]=0\, , \\
&i[H,S_2]=-\omega S_3\, ,\\
&i[H,S_3]=\omega S_2\, . 
\end{split}
\end{equation}
For the two-point operators, we obtain
\begin{equation}
\begin{split}
&i[H,S_1^2]=0\, , \\ &i[H,S_1S_2]=-\omega S_1 S_3\, ,\\ &i[H,S_1 S_3]=\omega S_1 S_2 \, , \\
&i[H,S_2^2]=-\omega \left\{S_3,S_2\right\}\, ,\\ &i[H,S_2 S_3]=\omega (S_2^2-S_3^2)\, ,\\
&i[H,S_3^2]=\omega \left\{S_2,S_3\right\}\, .
\end{split}
\end{equation}

Next, we want to look at the action of the dissipative contribution. Before doing that, in order to simplify the calculations, we note that 
$$
\mathbb{D}^*[O]=\frac{\Gamma}{N}\left(S_+ O S_--\frac{1}{2}\left\{S_+S_-,O\right\}\right)=\frac{\Gamma}{2N}\left([S_+,O]S_-+S_+[O,S_-]\right)\, , 
$$
and also that we can write 
$$
\mathbb{D}^*[O]=\frac{\Gamma}{2N}\left([[S_1,O],S_1]+[[S_2,O],S_2]\right)+i\frac{\Gamma}{2N}\left(\left\{[S_1,O],S_2\right\}-\left\{[S_2,O],S_1\right\}\right)\, .
$$
For the sake of later calculations, we note that  terms with double commutators give contributions which are subleading in the large-$N$ limit. In fact, the only relevant terms are those containing the anticommutators. Another important simplification comes from noticing that 
$$
\mathbb{D}^*[O_\alpha O_\beta]=O_\alpha \mathbb{D}^*[O_\beta ]+ \mathbb{D}^*[O_\alpha]O_\beta + \frac{\Gamma}{N}\left[S_+,O_\alpha\right]\left[O_\beta,S_-\right]\, ,
$$
and that the term involving the two commutators is subleading large-$N$ limit. 
This observation shows in fact that calculating these second-order moments is essentially equivalent to calculating derivatives of products of the mean-field variables, since the additional quantum correction terms arising from the Lindblad dissipator are subleading in the thermodynamic limit. This remark highlights the redundancy of the cumulant expansion in this regime, as the equations for the second-order moments do not contain additional information beyond the mean-field dynamics.

With the above simplifications, we can look at the action of the dissipator on the set of operators considered above. Accounting only for leading-order terms, we find 
\begin{equation}
\begin{split}
&\mathbb{D}^*[S_1]\approx \frac{\Gamma}{2N}\{S_3,S_1\}\, \\
&\mathbb{D}^*[S_2]\approx \frac{\Gamma}{2N}\{S_3,S_2\}\, ,\\
&\mathbb{D}^*[S_3]\approx -\frac{\Gamma}{N} S_1^2-\frac{\Gamma}{N} S_2^2\, .
\end{split}
\end{equation}
For the two-point operators, we obtain
\begin{equation*}
\begin{split}
&\mathbb{D}^*[S_1^2]\approx\frac{\Gamma}{2N}\left(\left\{S_3,S_1\right\}S_1+S_1\left\{S_3,S_1\right\}\right)\, , \\
&\mathbb{D}^*[S_2^2]\approx\frac{\Gamma}{2N}\left(\left\{S_3,S_2\right\}S_2+S_2\left\{S_3,S_2\right\}\right)\, \\
&\mathbb{D}^*[S_3^2]\approx-\frac{\Gamma}{N} (S_1^2S_3+S_2^2 S_3+S_3S_1^2 +S_3S_2^2) \, , \\
&\mathbb{D}^*[S_1 S_2]\approx\frac{\Gamma }{2N} S_1\{S_2,S_3\}+\frac{\Gamma}{2N}\{S_1,S_3\}S_2\, ,\\
&\mathbb{D}^*[S_1 S_3]\approx-\frac{\Gamma}{N} S_1(S_1^2 + S_2^2)+\frac{\Gamma}{2N} \{S_3,S_1\}S_3 \, , \\ &\mathbb{D}^*[S_2 S_3]\approx-\frac{\Gamma}{N} S_2(S_1^2+S_2^2)+\frac{\Gamma}{2N}\{S_3,S_2\}S_3\, .
\end{split}
\end{equation*}

Now, we define the moments $M_\alpha=\langle S_\alpha\rangle_t/N$, $M_{\alpha\beta}=\langle S_\alpha S_\beta\rangle_t/N^2$, and $M_{\alpha\beta\gamma}=\langle S_\alpha S_\beta S_\gamma\rangle_t/N^3$. Using the above relations, we find, in the thermodynamic limit, 
\begin{equation}
\begin{split}
    &\dot{M}_1= \Gamma M_{13}\, ,\\
    &\dot{M}_2=-\omega M_3+\Gamma M_{23}\, , \\
    &\dot{M}_3=\omega M_2-\Gamma (M_{11}+M_{22})\\
    &\dot{M}_{11}=2\Gamma M_{113}\, ,\\
    &\dot{M}_{12}=-\omega M_{13}+2\Gamma M_{123}\, ,\\
    &\dot{M}_{13}=\omega M_{12}-\Gamma (M_{111}+M_{122})+\Gamma M_{133}\, ,\\
    &\dot{M}_{22}=-2\omega M_{23} +2\Gamma M_{223}\, ,\\
    &\dot{M}_{23}=\omega (M_{22} -M_{33} )-\Gamma M_{112}-\Gamma M_{222}+\Gamma M_{233}\, ,\\
    & \dot{M}_{33}=2\omega M_{23}-2\Gamma (M_{113}+M_{223}) \, .
\end{split}
\end{equation}
To truncate these equations at second-order in the moments, one assumes that third-order cumulants are all vanishing so that one can write third-order moments in terms of lower-order ones. Namely, one uses, for vanishing cumulants, the following relation
$$
M_{\alpha\beta\gamma}=M_{\alpha \beta}M_\gamma +M_{\alpha\gamma} M_\beta + M_{\beta \gamma} M_\alpha- 2M_\alpha M_\beta M_\gamma\, .
$$

\begin{figure}[t]
    \centering
\includegraphics[width=0.6\columnwidth]{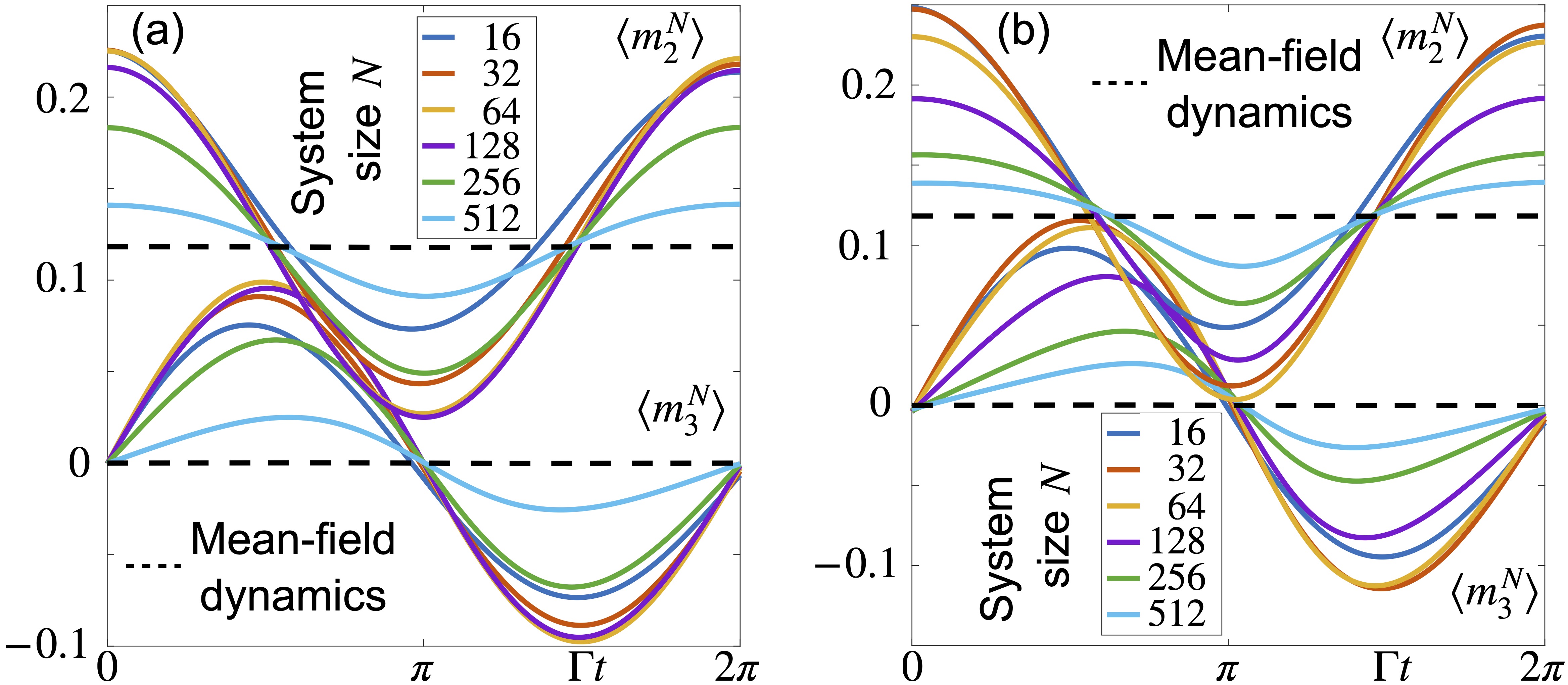}
    \caption{{\bf Time-invariant pure state.} (a) Numerically exact evolution of the expectation values $\langle m_2^N\rangle$ and $\langle m_3^N\rangle$ for the state in Eq.~\eqref{t-inv-btc-pure}. Here, we consider a spacing for the integral given by $ \Delta \tau_0 = 0.3$ (in units of $1/\Gamma$). The dashed line provides the prediction using the decomposed mean-field dynamics.  (b) Same as in (a) but for $ \Delta \tau_0 = 0.5$.  We consider here $\omega =\sqrt{5/4}$ (in units of $\Gamma$). }
\label{figS1}
\end{figure} 

\section*{V. Time-invariant pure state}
We consider here the superposition state 
\begin{equation}
\ket{\Psi}=\frac{1}{\sqrt{T}} \int_0^T {\rm d}\tau_0 \ket{\varphi_{\tau_0}}\, ,
\label{t-inv-btc-pure}
\end{equation}
where the states $\ket{\varphi_{{\tau_0}}}$ are clustering states such that 
$$
\bra{\varphi_{\tau_0}}m_2^N\ket{\varphi_{\tau_0}}=\frac{1}{2}\cos \theta(\tau_0)\, ,\qquad \bra{\varphi_{\tau_0}}m_3^N\ket{\varphi_{\tau_0}}=\frac{1}{2}\sin \theta(\tau_0) \, .
$$
The integral over $\tau_0$ spans a full period of the mean-field dynamics. To construct this state numerically, as also done for the state in Eq.~\eqref{t-inv-btc}, we have to discretize the integral through a finite spacing $\Delta \tau_0$. For the statistical mixture considered in the main text, we can chose an arbitrarily small $\Delta \tau_0$ (we have chosen $\Delta \tau_0=0.005$ in units of $1/\Gamma$) and we have observed a rapid convergence with $N$. On the other hand, when considering the pure state $\ket{\Psi}$ if we consider a small spacing $\Delta \tau_0$, we have the problem that $|\braket{\varphi_{\tau_0}|\varphi_{\tau_0+\Delta \tau_0}}|$ remains significantly different from zero up to system sizes which are larger than those we could reach numerically. As such, in order to see that also the pure state becomes time-invariant in the thermodynamic limit, we have to consider smaller spacings as we have done in Fig.~\ref{figS1}. 

\end{document}